\documentstyle[epsfig,  emulateapj5]{aastex}
\reversemarginpar
\makeatletter

\newenvironment{figurehere}
  {\def\@captype{figure}}
  {}
\makeatother
\begin{document}

\title{Probing the Intergalactic Medium with the \ion{O}{6} Forest}

\author{Taotao Fang AND Greg L. Bryan\footnote{Hubble Fellow}}
\affil{Department of Physics and Center for Space Research}
\affil{Massachusetts Institute of Technology}
\affil{NE80-6081, 77 Massachusetts Avenue, Cambridge, MA 02139}

\begin{abstract}

Recent STIS and FUSE observations have detected \ion{O}{6} absorption lines at low redshift that are not clearly associated with any galactic system. In this paper, we argue that these lines are due to metal enriched hot gas in the intergalactic medium.  Using numerical simulations of a cosmological-constant dominated cosmology, combined with reasonable assumptions about the metallicity distribution, we show that the number density and internal characteristics of these
lines are correctly predicted. We find that the \ion{O}{6} is primarily produced by collisional ionization from gas at a few times $10^{5}$ K for lines with equivalent widths larger than 40 m\AA, while weaker lines can also be produced by photo-ionization. The absorption occurs in diffuse gas in filaments at moderate overdensity ($\delta \sim 5-100$).

\end{abstract}

\keywords{intergalactic medium --- quasars: absorption lines --- large-scale structure of universe --- methods: numerical}

\section{Introduction}

The census of baryons in the local universe \citep{fhp98} indicates that a substantial fraction of the baryonic density predicted by primordial nucleosynthesis remains to be detected. Numerical simulations (\citealp{cos99a, dco01}) predict that a large fraction of this is in the form of a moderately warm/hot ($\sim 10^{5} < T < 10^{7}$ K) component (Warm-hot intergalactic medium, or WHIM).  The WHIM is hard to detect directly both because of the difficulty in UV observations as well as due to its relatively low intrinsic density. A promising direction is to look for its signature in absorption lines from background quasars (see \citealp{hgm98,plo98,fmb01}).

Recently, a number of groups have reported detecting the Li-like \ion{O}{6} resonance doublet in far-UV QSO spectra at $z<2$ (\citealp{ber94, bty96, lrr99, cch99}, rbh01). \citet{stl98} also detected an \ion{O}{6} absorber at $z=0.225$ without any \ion{C}{4} or \ion{N}{5} absorption, indicating that it was either collisionally ionized with a temperature $T \sim 5 \times 10^{5}$ or photo-ionized with a low density. \citet{tsa00} report an \ion{O}{6} absorption system at $z=0.14$, possibly associated with a galaxy group and \citet{tsj00} describe five \ion{O}{6} absorbers in the range $0.15 < z < 0.27$.  Adopting reasonable assumptions, these authors argued that the total number of baryons associated with these absorbers was significant, comparable to the combined mass in stars, cool gas and cluster gas.

In this paper, we explore the connection between these two ideas in more detail. In particular, we posit that the \ion{O}{6} absorption arises, in large part, in collisionally ionized gas that resides in the sheets, filaments and low density clumps that are naturally produced by gravitational collapse in hierarchical cosmology.  The gas is shock-heated to a range of temperatures, while \ion{O}{6} is only significantly produced in a relatively narrow range around $3 \times 10^5$ K.  This is near the peak of the cooling curve (e.g. \citealp{sdo93}) and so places a constraint on the density if the gas is not to cool in the Hubble timescale. For gas with $10\%$ solar metallicity at $3 \times 10^5$ K, the density must be less than $n_H \le 10^{-5}$ cm$^{-3}$. Assuming an \ion{O}{6} ionization fraction of 0.15, this requires a path length of 100 kpc to generate a column density of $N_{OVI} = 4 \times 10^{13}\ {\rm cm^{-2}}$, typical of observed values.  While this is a relatively large path length, the Hubble expansion if undisturbed would be only $\sim 7\ {\rm km\ s^{-1}}$, which is well within the observed line widths.  Therefore, we expect that thermal broadening should dominate and $b \sim 20\ {\rm km\ s^{-1}}$ for most lines.  As we will show; however, colder, photoionized gas can also give rise to \ion{O}{6} absorption.  For these lines, which tend to have lower column densities, the predicted line width is considerably smaller. Very recently, a preprint by \citet{cto01} has appeared which examines this issue using different simulation techniques; where there is overlap, our results agree reasonably well.

This paper is organized as follows. In section~2 we describe the hydrodynamic simulation and how we synthesis the simulated spectra. Section~3 discusses the results and Section~3 summarizes our main conclusions.

\section{Method \label{sec:method}}

We use numerical simulations to generate predictions for the observed \ion{O}{6} distribution.  In particular, we use a grid-based Adaptive Mesh Refinement (AMR) method (\citealp{bry96, nbr99}) that provides both good shock capturing as well as relatively high resolution in dense regions. We simulate a cube 20 $h^{-1}$ Mpc on a side with gas and dark matter mass resolution of $6 \times 10^7$ and $5 \times 10^8 M_{\odot}$, respectively.  The smallest grid cell (i.e. the best resolution) is 9.8 kpc and occurs in the densest regions.  We assume a cosmological model with a matter density $\Omega_0 = 0.3$, baryon density $\Omega_b = 0.04$, Hubble constant $h=0.67$ (in units of $100\ {\rm km\ s^{-1}Mpc^{-1}}$) and cosmological constant $\Omega_{\Lambda} = 0.7$. The initial density field is drawn from an adiabatic CDM power spectrum as approximated by the formula of \citet{ehu98}.

We include dark matter and gravity but not radiative cooling or energy/metal injection from SN-driven galactic winds. By comparing a set of cosmological hydrodynamic simulations, \citet{dco01} concluded that radiative cooling play a minor role in the evolution of WHIM gas at $z \lesssim 3$ due to its low density. However, these processes can play an important role in metal enrichment of the intergalactic medium (IGM). For instance, {\sl FUSE} has detected \ion{O}{6} absorption in the galactic wind from the prototypical dwarf galaxies (see, e.g., \citealp{hsm01}). Since we do not know the distribution of oxygen a priori in this simulation, we adopt a density dependent metallicity model as derived in the simulations of \citet{cos99b}. While this is reasonable, we plan to revisit this issue in a future paper using a simulation which self-consistently includes both radiative cooling and metal/energy injection.

To analyze the results, we generate artificial spectra as described in \citet{fbc01} and \citet{zan97}, both for the 1032 $\AA$ \ion{O}{6} line and the 1216 $\AA$ HI Ly$\alpha$ line.  To compute the $f(OVI/O)$ and $f(HI/H)$ fractions, we adopt two models to investigate the effect of collisional ionization and photo-ionization mechanisms. In model A, we use collisional ionization only, with ionization fractions from \citet{mmc98}. In model B, we use Cloudy (version 90.04; \citealp{fkv98}) to generate a grid of temperatures ($10^{3} < T < 10^{7}$ K) and densities ($10^{-8} < n ({\rm cm^{-3}}) < 10^{-2}$). We included a background radiation field as computed by \citet{hma96} at $z=0$ due to the observed quasar distribution. We adopt a mean specific intensity at the Lyman limit of $J_{\nu} = 2\times10^{-23}\ {\rm ergs\ s^{-1}Hz^{-1}sr^{-1}}$. Since the simulation did not include radiative heating/cooling, the unshocked IGM gas has a temperature which is unrealistically low. For these low temperature (and generally low density) regions, we adopt the relation between temperature and density that has been found in simulations of the Ly$\alpha$ forest: $T = T_{0}(1+\delta)^{\gamma-1}$ where $T_{0} \sim 5,000$ K, and $\gamma = 1.4$ (see, e.g., \citealp{zan97, rgs00, dtr01} for observational support). Here overdensity $\delta = \rho_{b}/\left<\rho_{b}\right> - 1$ and $\left<\rho_{b}\right>$ is the mean baryon density of the universe. To fit the simulated spectra we adopt the method developed by \citet{zan97}, which should work well for these relatively isolated lines.

\section{Results\label{sec:results}}

Figure~1 displays the projected baryon overdensity (panel a), \ion{O}{6} column density (panel b, based on model B), temperature ($10^{4}-10^{7.5}$ K in panel c; $10^{5}-10^{6}$ K in panel d). Comparing panel b with panel a, we find that most of the high column density \ion{O}{6} lines come from regions with overdensities of $\delta \sim 5-100$.  Collisional ionization is dominant in many of these regions, given their temperatures in the range $10^{5}-10^{6}$ K (panel d), but some systems are clearly photo-ionized, particularly at the low density end. Note that big clusters and groups are too hot to produce \ion{O}{6}.

\begin{figurehere}
\centerline{\psfig{file=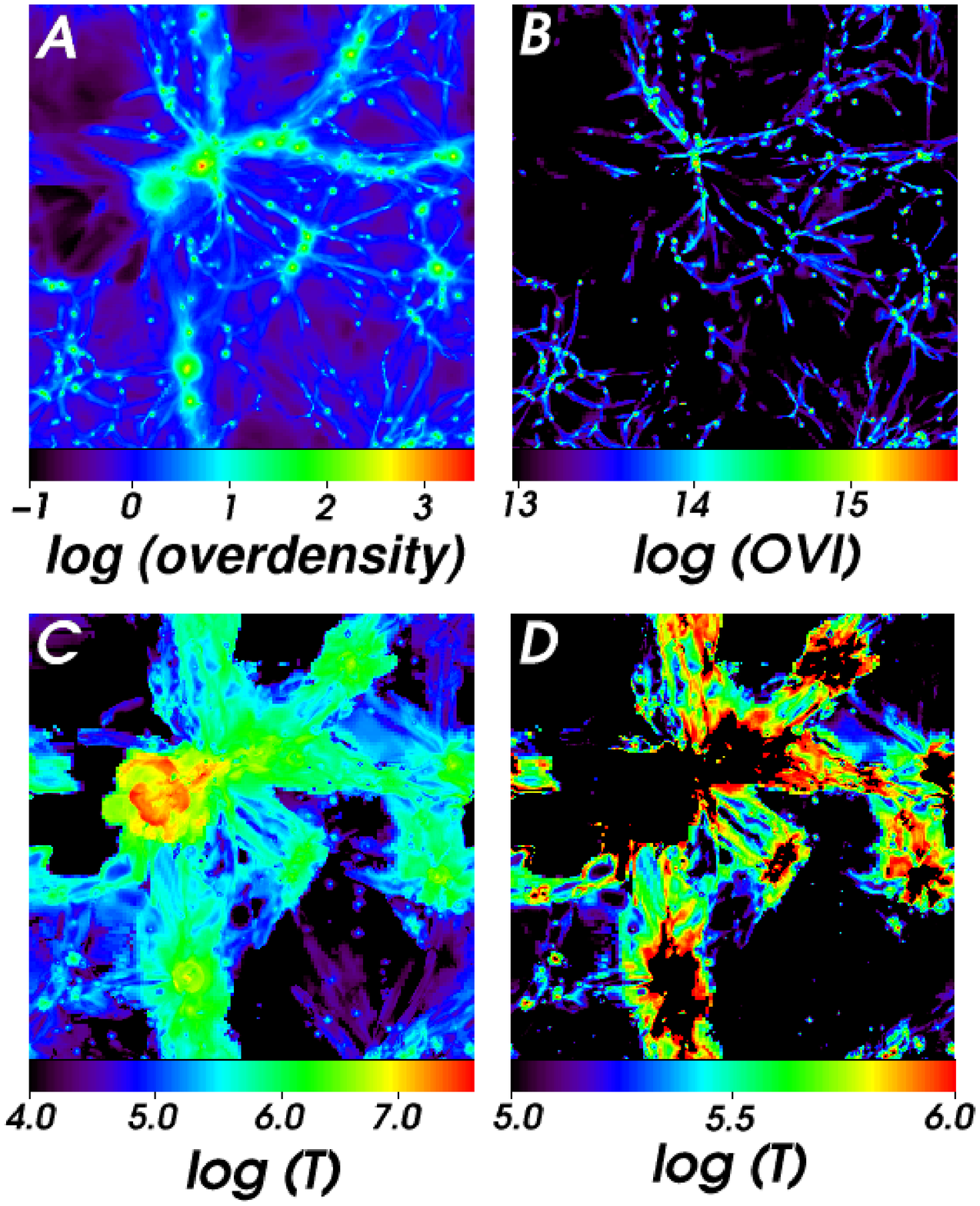,width=9cm}}
\caption{Here we show various quantities from the simulation projected along a 5 $h^{
-1}$Mpc line-of-sight; the images are 20 $h^{-1}$ Mpc on a side.  Panel a shows the m
ean baryon overdensity, \ion{O}{6} column density (panel b, in units of ${\rm cm^{-2}
}$), temperature ($10^{4}-10^{7.5}$ K in panel c; $10^{5}-10^{6}$ K in panel d). Most
 high column density \ion{O}{6} comes from regions with overdensities of $\delta \sim
 5-100$ (panel a) and temperatures between $10^{5}-10^{6}$ K (panel d).\label{fang_fi
g1}}
\end{figurehere}
\vspace{0.2cm}

In Figure~2 we show the properties of the simulated region along one random line-of-sight (LOS). From top to bottom we display the baryon number density, peculiar velocity, temperature, \ion{O}{6} number density and transmission spectrum ($F = \exp(-\tau)$, where $\tau$ is the optical depth) for \ion{O}{6} in model A (solid line), model B (dashed line) and for \ion{H}{1} (dot-dashed line). Model B has been shifted by $5\%$ for visual clarity.  Note that at high densities and warm temperatures, both models agree well, while for low temperature regions, model B (which includes both photo- and collisional-ionization), produces substantially more \ion{O}{6}.

Figure~2e shows that both models generate a strong \ion{O}{6} absorption line at around $z\sim 0.0019$ along this LOS.  Notice that no strong peak appears in the $n$ and $n_{OVI}$ distributions at the corresponding redshift. This absorption line is actually produced by the $n_{OVI}$ peak at $z\sim 0.0044$. It is the peculiar velocity which shifts the line to a lower redshift: a peculiar velocity of $\sim 750\ {\rm km\ s^{-1}}$ corresponds to a shift of $\Delta z \sim 0.0025$.  Parenthetically, we note that this means the total number of lines may be underestimated because the peculiar velocities can shift lines out of the simulated regions; however, we estimate that this would only change our predicted number density by about $10\%$. Fitting the observed line in Figure~2e gives an equivalent width (EW) of $\sim 185\ {\rm m\AA}$ and a Doppler $b$-parameter of $\sim 21\ {\rm km\ s^{-1}}$, which is roughly consistent with thermal broadening ($b = 0.129(T/A)^{1/2}\ {\rm km\ s^{-1}}$, $A$ is the atomic mass number).

\citet{tsa00} suggested that a broad, hot \ion{H}{1} Ly$\alpha$ component may coexist with the \ion{O}{6} absorbers and such a broad component could be easy to miss with current observations. To demonstrate this, they showed that a fit to the multicomponent Lya profile including a broad component (due to \ion{H}{1} coexisting with the \ion{O}{6} in hot gas) is indistinguishable from a fit which does not contain such a broad component, at the signal-to-noise level of their data. In Figure~2e we display the \ion{H}{1} Ly$\alpha$ spectrum (the dot-dashed line) from our simulation. A broad component is clearly identified at the position of the strongest \ion{O}{6} absorption line. Spectral fitting shows this line has a column density of $3.2 \times 10^{13}\ {\rm cm^{-2}}$ and a Doppler $b$-parameter of $\sim 76\ {\rm km\ s^{-1}}$. Both parameters are consistent with those from the missing \ion{H}{1} Ly$\alpha$ absorption line in \citet{tsa00}. The Doppler $b$-paramter is also consistent with thermal broadening by a hot gas of $\sim 3-4 \times 10^{5}$ K. \footnote{A recent observation of the low redshift quasar E~1821+643 \citep{tgs01} reveals that in additional to a narrow \ion{H}{1} Ly$\alpha$ line, a relatively broaden component appears with $b \approx 85\ {\rm km\ s^{-1}}$ in the $z = 0.1212$ \ion{O}{6} absorption systems.}

\begin{figurehere}
\centerline{\psfig{file=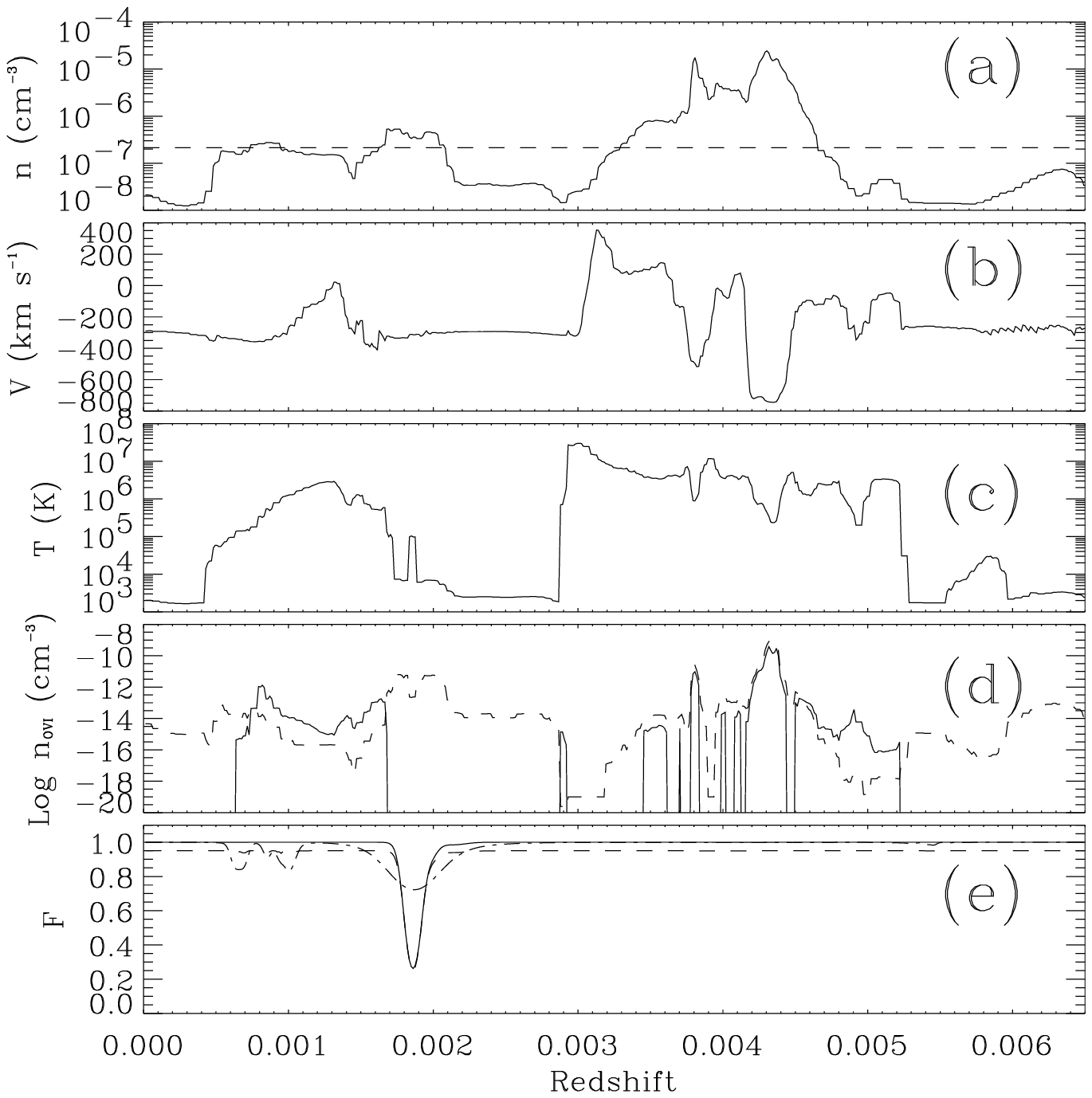,width=9cm}}
\caption{Physical properties along one random LOS. (a) baryon number density; (b) pec
uliar velocity; (c) temperature; (d) \ion{O}{6} number density (model A: solid line; 
model B: dashed line); (e) synthesized spectra for \ion{O}{6} and \ion{H}{1} (dot-das
hed line). In panel (d) and (e) model B is shifted down by $5\%$ for visual clarity. 
The dashed line in panel (a) indicates the mean baryon density. The lines at $z\sim 0
.0019$ are produced by gas at $z\sim 0.0044$.\label{fang_fig2.eps}}
\end{figurehere}
\vspace{0.2cm}

By applying the spectral fit procedure to a total of $3,000$ LOS we obtain the cumulative number per unit redshift ($dN/dz$) as a function of \ion{O}{6} equivalent width. Figure~3 displays the $dN/dz$ for both models. We also plot two observational results from \citet{tsa00} for PG 0953+415 and \citet{tsj00} for E~1821+643. In general we find that the simulation and observations agree reasonably well. We notice that for $EW > 40\ {\rm m\AA}$, the \ion{O}{6} absorption line number per unit redshift from model B is less than twice the line number from model A. This implies that above $EW = 40\ {\rm m\AA}$ the lines from collisionally ionized gas begin to outnumber lines from photo-ionized gas. At $EW > 80\ {\rm m\AA}$, both models give consistent distributions. This means that those strong lines come from high density and high temperature regions, where collisional ionization dominates. For $EW < 40\ {\rm m\AA}$, lines from photo-ionized gas outnumber lines from collisionally ionized gas, implying that lines produced by photo-ionized gas contribute significantly for low strength lines. Due to our simulation resolution lines with $EW < 20\ {\rm m\AA}$ could be underestimated. Following \citet{rtr00} and \citet{tsa00} we estimate that at $EW > 40\ {\rm m\AA}$ where collisional ionization dominates, the gas probed by \ion{O}{6} absorption lines is about $10\%$ of the total baryonic matter. This means that the \ion{O}{6} absorption lines can actually probe about $\sim 30\%$ of the WHIM gas, since the fraction of the baryons predicted to be in the WHIM gas is about one third of the total baryonic matter \citep{dco01}. The remaining $\sim 70\%$ WHIM gas is likely to be too hot to be detected via \ion{O}{6} absorption: \citet{dco01} predicted that at $z=0$ the baryon fraction peaks at around $4\times 10^{6}$ K, in which \ion{O}{7} or \ion{O}{8} should be dominated.

\begin{figurehere}
\centerline{\psfig{file=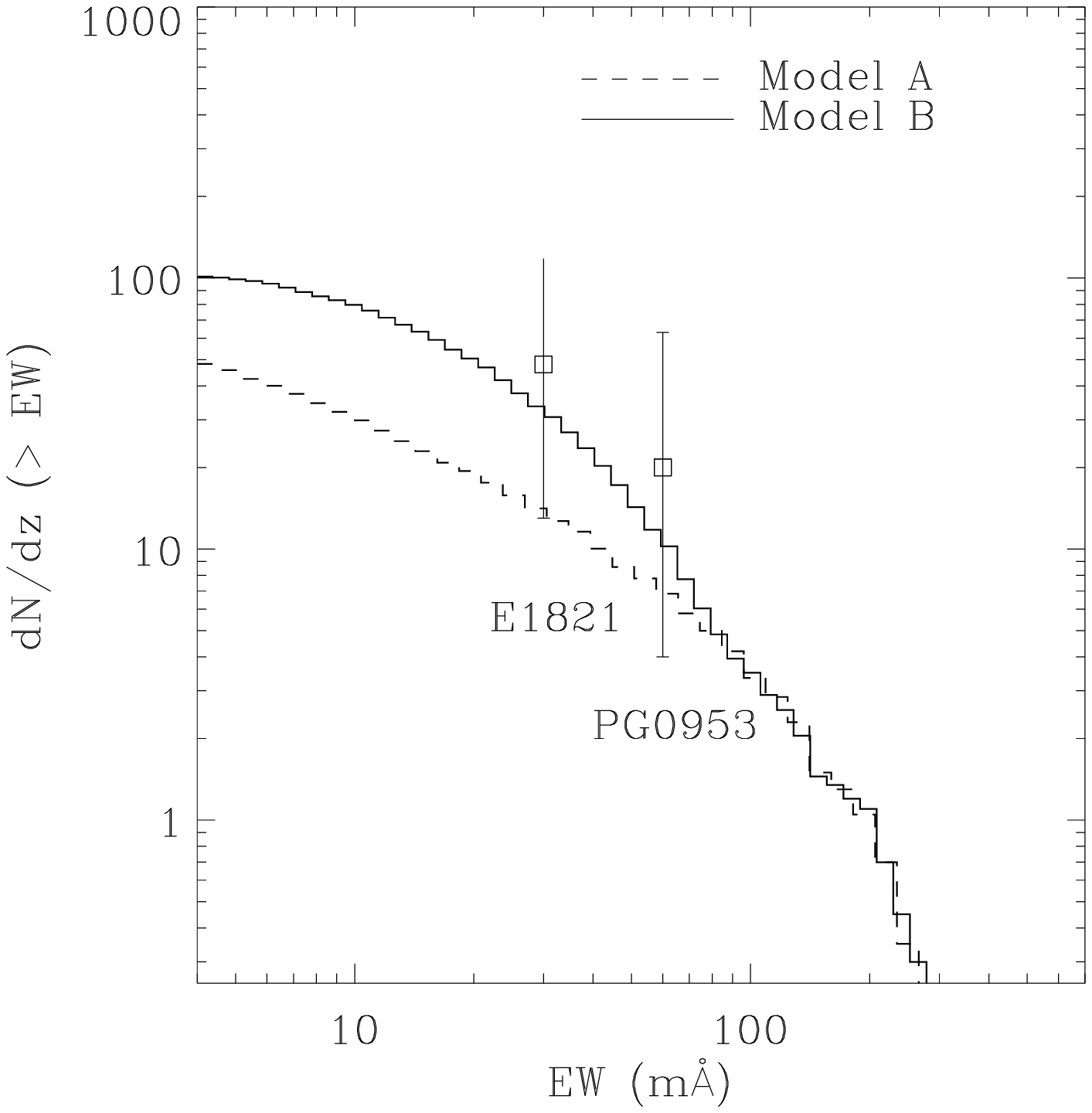,height=6cm,width=7cm}}
\caption{Cumulative number of \ion{O}{6} absorption lines per unit redshift vs. line 
equivalent width. Dashed line: model A; solid line: model B. Lines produced by collis
ionally ionzed gas dominate at $EW > 40\ {\rm m\AA}$. Observational data are plotted 
with $1\sigma$ error bars (\citealp{tsa00, tsj00}).\label{fang_fig3}}
\end{figurehere}
\vspace{0.2cm}

Our result is consistent with what \citet{cto01} find in their simulations, especially in that they find at $EW > 35\ {\rm m\AA}$ lines from collisionally ionized gas outnumber lines from photo-ionized gas. However, overall we predict more \ion{O}{6} absorption lines.

We find the Doppler $b$ parameter of the \ion{O}{6} absorption line can provide important clues in distinguishing photo- and collisional-ionization. By fitting the spectra, the Doppler $b$ parameters are obtained for all the \ion{O}{6} absorption lines. We find that the distribution of the line widths is well approximated by a log-normal: $f \propto \exp \left[-\log^{2}(b/b_{ln})/2\sigma_{ln}^{2}\right]$. In figure~4, we plot $b_{ln}$ against equivalent width cutoff for our two models. In the collisional ionization case (model A), the distribution is nearly constant with $b_{ln} \sim 20 {\rm km/s}$. On the other hand, when we include a photo-ionizing background (model B), the distribution depends on the strength of the line. For lines with $EW > 70 {\rm m\AA}$, the value of $b_{ln}$ is still around 20 km/s, but drops quickly for smaller lines. This reflects the fact that more photo-ionized lines contribute for $EW < 70 {\rm m\AA}$. 

\begin{figurehere}
\centerline{\psfig{file=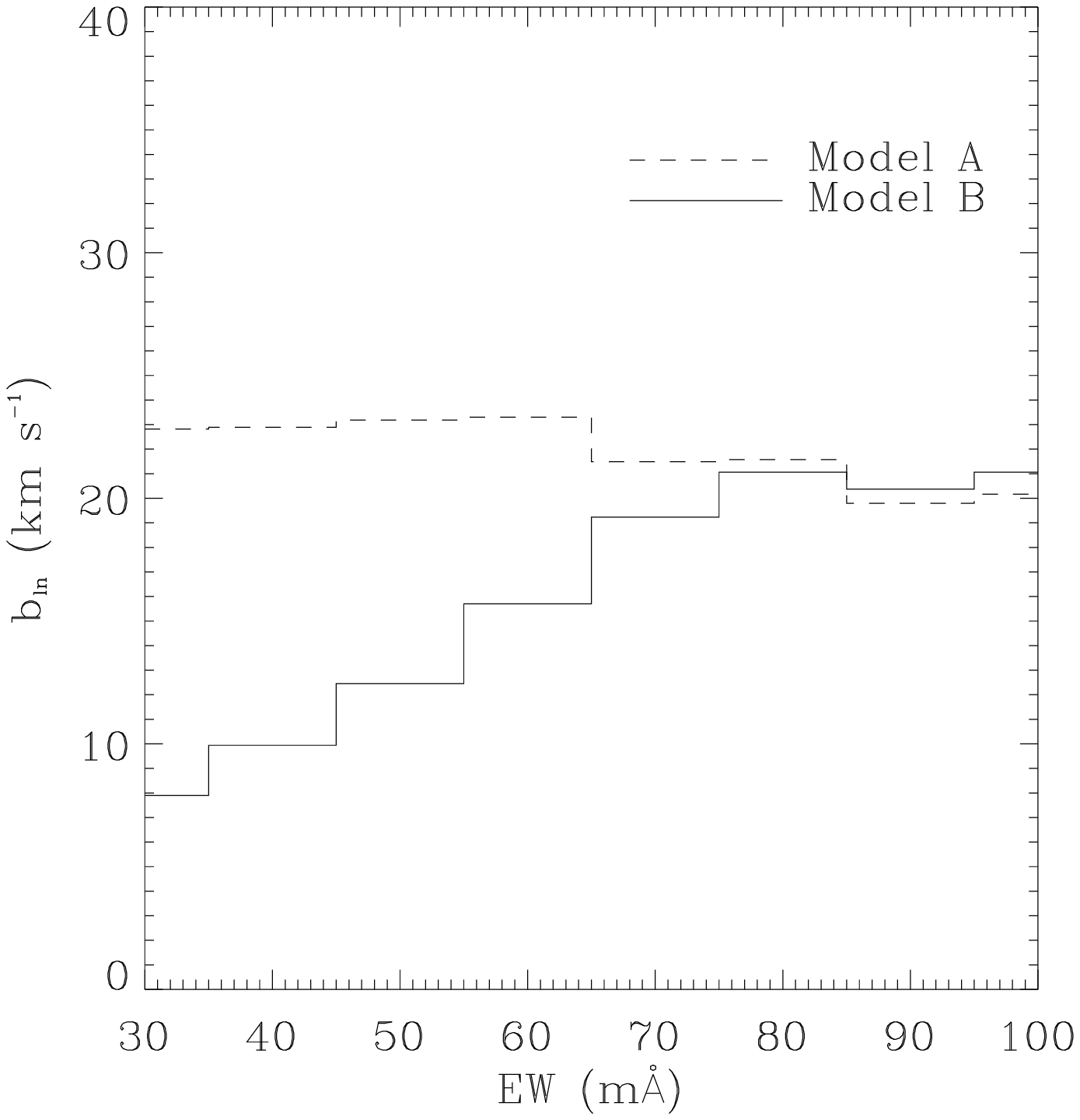,height=6cm,width=7cm}}
\caption{The peak of the Doppler $b$ distribution ($b_{ln}$) for lines with a EW larg
er than the corresponding value. The value of $b_{ln}$ is obtained by fitting each di
stribution with a lognormal functional form (solid line: model A; dashed line: model 
B). The \ion{O}{6} absorption lines that are produced by collisionally ionized gas ha
ve a mean $b$ value of $\sim 20\ {\rm km\ a^{-1}}$. After including photo-ionized gas
, the characteristic $b$ value drops as low as $\sim 5\ {\rm km\ s^{-1}}$.\label{fang
_fig4}}
\end{figurehere}
\vspace{0.2cm}

A statistically complete sample does not yet exist to compare against these predictions. Recently \citet{tri01} compiled the first batch of $\sim 15$ \ion{O}{6} absorption lines detected in a total of five quasars. They found that the majority of the \ion{O}{6} lines have $b$-values of $\sim 22\ {\rm km\ s^{-1}}$, implying a collisional-ionization origin. If this result continues to hold for low equivalent width lines, it may be telling us something either about the strength of the photoionizing background or the metallicity of the low-temperature filaments. Either case (lower background or lower metallicity in the cold gas), would produce fewer photo-ionized lines relative to collisionally ionized.

\section{Conclusions\label{sec:conclusions}}

In this {\it Letter} we have studied the properties of \ion{O}{6} absorption lines via hydrodynamic simulation.  Our main conclusions are as follows.

\begin{enumerate}

\item The projected \ion{O}{6} number density distribution shows that the majority of \ion{O}{6} absorption comes from filaments that connect virialized intersections or groups of galaxies. These regions typically have an overdensity of $\sim 5-100$.

\item Spectral fits show that collisional ionization dominates in high density, high temperature regions, while photo-ionization dominates in low density, low temperature regions.

\item The predicted cumulative distribution of \ion{O}{6} absorption lines fits the observations remarkably well, given the uncertainties due to limited resolution and the assumed metallicity distribution. At $EW > 40\ {\rm m\AA}$ \ion{O}{6} absorption lines produced by collisionally ionized gas start to outnumber lines produced by photo-ionized gas, while at $EW > 80\ {\rm m\AA}$ essentially all the \ion{O}{6} absorption lines are produced by collisionally ionized gas.

\item We find that the \ion{O}{6} absorption lines produced by collisionally ionized gas have a characteristic Doppler $b$ parameter of $\sim 20-23\ {\rm km\ s^{-1}}$, and lines due to the photo-ionized gas typically have narrower $b$-parameters.

\end{enumerate}

\acknowledgments{TF thanks the MIT/CXC team for support. This work is supported in part by contracts NAS 8-38249 and SAO SV1-61010. Support for GLB was provided by NASA through Hubble Fellowship grant HF-01104.01-98A from the Space Telescope Science Institute, which is operated by the Association of Universities for Research in Astronomy, Inc., under NASA contract NAS 6-26555.}

\end{document}